\documentclass{ws-p8-50x6-00}
\newcommand{\abs}[1]{\left| #1\right|}
\renewcommand{\Re}[1]{\mathop{\mathrm{Re}}\left\{ #1 \right\}}
\newcommand{\ltap}[1]{\ \raisebox{-.4ex}{\rlap{$\sim$}} \raisebox{.4ex}{$<$}\ }
\newcommand{\eq}[1]{eq.~(\ref{#1})}
\begin{document}

\title{``Secret'' neutrino interactions\footnote{
{\sc Neutrino Mixing} (Torino, March 1999) 
In honour of {\em Samoil Bilenky}'s 70-th Birthday}
}

\author{Mikhail Bilenky}

\address{Institute of Physics, AS CR, 18040 Prague 8, and \\
Nuclear Physics Institute, AS CR, 25068 Rez (Prague), Czech Republic}

\author{Arcadi Santamaria}

\address{Departament de F\'{\i}sica Te\`orica,
        IFIC, Universitat de Val\`encia--CSIC,\\
        E-46100 Burjassot (Val\`encia) Spain}  


\maketitle

\abstracts{
We review the information about a potentially strong 
non-standard four-neutrino interaction that can be obtained from available
experimental data. By using LEP results and nucleosynthesis
data we find that a contact four-fermion neutrino interaction that
involve only left-handed neutrinos or both left-handed and
right-handed neutrinos cannot be stronger than the standard weak
interactions. A much stronger interaction involving only
right-handed neutrinos is still allowed.}

\section{``Secret'' neutrino interactions?}

In the standard model (SM) the interaction between
neutrinos is given by the exchange of the $Z$-boson and the effective
Hamiltonian of the neutrino-neutrino interaction has the form
\begin{equation}
{\cal H}_{SM}^{\nu-\nu} = \frac{G_F}{\sqrt{2}}\sum_{\ell,\ell'=e,\mu,\tau}
(\bar{\nu}_{\ell L} \gamma_\alpha \nu_{\ell L})
(\bar{\nu}_{\ell' L} \gamma^\alpha \nu_{\ell' L})  \ ,
\label{standardnn}
\end{equation}
where $G_F$ is the Fermi constant. However, it is very difficult to
perform direct experimental tests on this interaction.

Many years ago the question was raised\cite{bile1,bbnc}
whether an additional non-standard four-neutrino interaction exists:
\begin{equation}
{\cal H}_I^{\nu-\nu} = F(\bar{\nu}\gamma_\alpha \nu)
(\bar{\nu} \gamma^\alpha \nu)
\label{effham}
\end{equation}
Such an effective interaction could arise, for instance, 
from the exchange of a strongly
interacting heavy vector boson, $V_{\mu}$, coupled to neutrinos only
\begin{equation}
{\cal L} = g_V (\bar\nu_i \gamma_\mu \nu_i) V^\mu~,
\end{equation}
when it is considered at energy scales much lower than the vector boson
mass, $m_V$, e.i. $q^2 \ll m_V^2$. In this case we have the relation
$F=g_V^2/m_V^2$.

Interactions mediated by scalars can
also be written in the form of ${\cal H}_I^{\nu-\nu}$ after a Fierz
transformation. The flavour structure could be, however, more general.

Obviously, the possible
effect of the exchange of a light particle coupled to neutrinos
can not be approximated correctly by the contact four-neutrino interaction
of the form (\ref{effham}) at $q^2 \ge m_V^2$. However, in this case 
the non-standard particles could be produced directly and such models 
are in general  
strongly constrained.
This is what happens for a class of popular models in which the
non-standard interaction among neutrinos is due to the exchange of a massless
majoron\cite{singlet,noytriplet,triplet,doublet,susymajoron}.
In this case very strong bounds on neutrino-majoron coupling constants
$g_{\ell\ell}$ follow from  searches for massive
neutrinos and neutral particles in $K\rightarrow \ell+\cdots$. One obtains 
\cite{majbounds}
$g_{ee}^2 < 1.8\cdot 10^{-4}$, $g_{\mu\mu}^2 < 2.4\cdot 10^{-4}$.
Consequently
majoron bremsstrahlung from neutrinos can give only a small
contribution to the invisible $Z$-width.  Majorons with non-vanishing
hypercharge\cite{triplet,doublet,susymajoron} could potentially
give a large contribution to the invisible $Z$-width because the $Z$-boson
can decay directly to scalars. Therefore, models of this type
have already been excluded by LEP data (in the case of the triplet majoron 
with hypercharge one\cite{triplet}
this contribution is equivalent to the existence of two additional
neutrinos and in the case of the doublet majorons
\cite{doublet,susymajoron} a contribution, equivalent to half the contribution 
of an additional neutrino, arises). Singlet majorons\cite{singlet} or 
non-singlet majorons without hypercharge\cite{noytriplet} cannot be 
excluded by LEP data.

We review in this paper, from a historical point of view, the 
different bounds set on the exotic four-neutrino contact interaction. 
In section \ref{sec:lowenergy}
we review the old low-energy bounds. In section \ref{sec:sn} we take a look
to the limits obtained from the supernova SN1987A. Sections \ref{sec:4nus}
and \ref{sec:2nus} are devoted to the limits obtained from the invisible
decay width of the $Z$ gauge boson measured at LEP: section \ref{sec:4nus}
by using the four-neutrino decay at tree level and section \ref{sec:2nus}
by using the one loop contribution of the SNI to the two-neutrino decay.
In section \ref{sec:wdecay} we briefly discuss the possibility of gaining
some information on the SNI by modification of the lepton spectra in 
the $W$-boson decay due to the process 
$W^+\rightarrow \ell^+\nu_\ell\nu\bar{\nu}$. In section \ref{sec:ns} we
present the stringent bounds obtained from nucleosynthesis and finally
in section \ref{sec:conclusions} we present our conclusions.

\section{Low-energy bounds}
\label{sec:lowenergy}

In the pioneering paper \cite{bile1}
the contributions of  ${\cal H}_I^{\nu-\nu}$ to different low-energy 
processes were analysed and different bounds were set.

The SNI, contributes to the decays
$\pi^+ \rightarrow e^+ \nu_e \bar{\nu} \nu$ and
$K^+ \rightarrow l^+ \nu_l \bar{\nu} \nu ~(l=e,\mu)$,
and could modify the inclusive lepton energy spectra
in $K^+$ and $\pi^+$ decays, which are dominated by standard
decays, $K^+ (\pi^+)\rightarrow l^+ \nu_l$.
From an analysis of these
spectra the following bounds on the coupling $F$ were 
obtained~\cite{bile1}
\begin{equation}
|F| \le 10^7 G_F,~~~~~~|F| \le 2\times 10^6 G_F~,
\label{firstbounds}
\end{equation}
where $G_F$ denotes the weak Fermi constant.
\par

Similar bounds were found \cite{bile1} from the absence of leptons
with ``wrong'' charge in the process
$\nu_{\mu}~+~N \rightarrow \mu^+~+~ \nu_\mu~+~ \nu_\mu~+X$.
\par
Later on these bounds were improved in a special 
experiment \cite{CHP} (for more recent experiments see 
also \cite{final})  
searching for the decay
$K^+ \rightarrow \mu^+ \nu_{\mu} \bar{\nu} \nu $.
From the negative result of this experiment the following limit was set:
\begin{equation}
F \le 1.7\times 10^5 G_F~.
\label{nextbounds}
\end{equation}
\par

It seemed at that time that four-neutrino interactions 
could be much stronger than standard model neutral current interactions. 

The reason why bounds on the non-standard neutrino interaction coming
from low-energy experiments are so loose is evident.
The SNI contributes only to the decays
with four particles in the final state, and such processes
are strongly suppressed by phase space compared with the standard
leptonic $\pi$ and $K$ decays.

\section{Supernova bounds}
\label{sec:sn}

The detection of (anti)neutrinos from {SN1987A} stimulated again the 
interest on SNI. Using that data some new limits were set.

In particular, the detection of neutrinos from {SN1987A} 
requires that the value of the mean
free path of neutrinos through the cosmic background particles (CBP) is 
comparable or greater than the distance to the supernova. 

Stable neutrinos
should be present today as CBP, therefore, a four-neutrino interaction
will contribute to the mean free path of supernova neutrinos and a bound
can be set\cite{sn87back}.

If neutrinos have an interaction with neutrinos mediated by heavy vector 
bosons with mass $M$ and coupling $g$, a bound on $g/M$ was obtained
\begin{equation}
\frac{g}{M} < \frac{12}{MeV}\ ,
\end{equation}
which can be translated into the following bound on the constant $F$:
\begin{equation}
{F} < {10^{13}}\, G_F\ .
\end{equation}
This bound is much worse than the obtained low-energy bounds.

On the other hand, from the estimate of the diffusion time
of neutrinos in the supernova and its comparison with the duration of the
detected neutrino pulse, an upper bound on
neutrino-neutrino cross section $\sigma_{\nu\nu}$ can 
be obtained\cite{sn87diff} (see also \cite{Dicus:1989jh})
\begin{equation}
\sigma_{\nu-\nu} < 10^{-35}\, cm^2\ .
\label{difftime}
\end{equation}
If this cross section arises from a strong four-neutrino interaction
one can obtain the following estimate
\begin{equation}
{F}< {10^3}\, G_F\ ,
\end{equation}
which is two orders of magnitude better than the best of the low-energy
bounds but still allows for a rather strong SNI.

\section{The decay $Z\rightarrow \nu \bar\nu \nu \bar\nu$}
\label{sec:4nus}

The decays $\pi (K) \rightarrow \ell \nu_\ell \nu \bar{\nu}$  with
four particles in the final state are strongly suppressed by phase
space in comparison with the usual two-body lepton decays. 
Decays of much heavier
particles, such as gauge bosons, will provide a much larger phase space for
multi-neutrino production.

If strong four-neutrino interactions exist, four-neutrino decays
\begin{equation}
Z\rightarrow \nu \bar{\nu} \nu \bar{\nu}
\label{fournudecay}
\end{equation}
will also contribute to the invisible width of the $Z$ gauge boson
and therefore the strength of such interaction can be 
constrained\cite{BBS93} from the precise LEP measurement of 
$\Gamma_{invis}$.

\begin{figure}[t]
\begin{center}
\epsfxsize=20pc 
\epsfbox{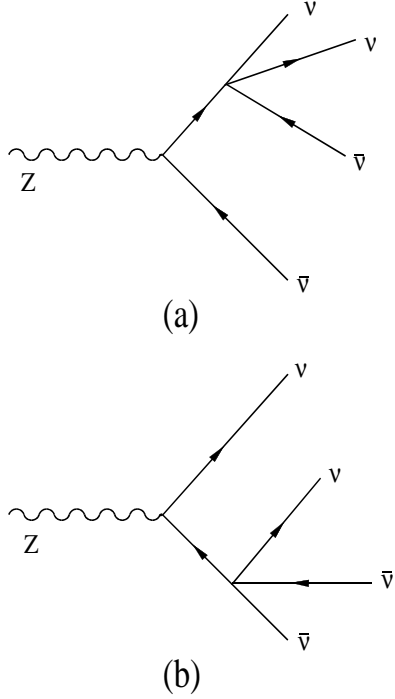}
\end{center} 
\caption{Diagrams contributing to the process
$Z\rightarrow \nu\bar{\nu}\nu\bar{\nu}$.\label{fig:fig1}}
\end{figure}

To be definite we take the Hamiltonian of $\nu-\nu$ interactions with
a general $V,A$ form
\begin{equation}
{\cal H}_I^{\nu-\nu} = {F} \sum_{\ell,\ell'=e,\mu,\tau}
(\bar{\nu}_{\ell} {O_{\ell\alpha}} \nu_{\ell})
(\bar{\nu}_{\ell'} {O^\alpha_{\ell'}} \nu_{\ell'})\ .
\label{inter}
\end{equation}
Here
\begin{equation}
{O_\ell^\alpha}= {a_\ell} \gamma^\alpha P_L
+ {b_\ell} \gamma^\alpha P_R\ ,
\end{equation}
$P_L$ and $P_R$ are the left and right chirality projectors
$P_L=\frac{1}{2}(1-\gamma_5)$, $P_R=\frac{1}{2}(1+\gamma_5)$ and
$F,a_\ell,b_\ell$ are real parameters.

For the total probability of the decay of $Z$-bosons into two
neutrino pairs (identical and non-identical) we have found the
following expression~\cite{BBS93}
\begin{equation}
\Gamma(Z\rightarrow \nu\bar{\nu}\nu\bar{\nu}) =
\frac{G_F}{\sqrt{2}} m_Z^7 F^2
\frac{1}{1024\pi^5} \sum_{\ell,\ell'} \left\{
(a_\ell^2 a_{\ell'}^2+a_\ell^4 \delta_{\ell\ell'}) C_1+
a_\ell^2 b_{\ell'}^2 C_2 \right\}\ ,
\label{widths}
\end{equation}
where
\begin{equation}
C_1=-\frac{1651}{486}+\frac{28}{81}\pi^2, \hspace{1.2cm}
C_2=\frac{259}{486}-\frac{4}{81}\pi^2
\end{equation}

The summation runs over the three generations $\ell,\ell'=e,\mu,\tau$.

The decays $Z\rightarrow \nu\bar{\nu}\nu\bar{\nu}$ are 
not sensitive to pure right-handed $\nu-\nu$ interactions. 
This is also true for any
process involving neutrinos produced through the standard interaction.

Assuming {$e-\mu-\tau$ universality} in the
non-standard $\nu-\nu$ interaction ($a_\ell=a\ ,b_\ell=b,\ \ell=e,\mu,\tau$),
one can rewrite the expression for the decay width of
$Z\rightarrow \nu\bar{\nu}\nu\bar{\nu}$ in the following form :
\begin{equation}
\Gamma(Z\rightarrow \nu\bar{\nu}\nu\bar{\nu})=
\frac{G_F}{\sqrt{2}} {m_Z^7} \frac{1}{1024\pi^5}
{\bar{F}^2} (12 C_1+9{\bar{b}^2} C_2)  \ .
\label{parwidth}
\end{equation}
Here {$\bar{F}^2\equiv F^2 a^4$} and the parameter 
{$\bar{b}^2\equiv b^2/a^2$}
characterises the relative contribution of right-handed currents into
the $\nu-\nu$ interaction. 

Assuming that only the standard decays 
$Z\rightarrow \nu_\ell \bar{\nu}_\ell\ \ (\ell=e,\mu,\tau)$
and the decays $Z\rightarrow \nu\bar{\nu}\nu\bar{\nu}$
contribute to the invisible width of the $Z$-boson $\Gamma_{invis}$
we can obtain a bound on $\bar{F}$.
\begin{equation}
\Gamma_{invis}=3\Gamma(Z\rightarrow \nu_\ell \bar{\nu}_\ell)^{SM}+
\Delta \Gamma_{invis}\ .
\label{invisible}
\end{equation}

In our case, 
\begin{equation}
\Delta \Gamma_{invis} = \Gamma(Z\rightarrow \nu\bar{\nu}\nu\bar{\nu})
\label{deltainv}
\end{equation}
On the other hand this quantity can also be expressed as
\begin{equation}
\Delta \Gamma_{invis}=
\Gamma_{invis} -
3 \left(\frac{\Gamma_{\bar{\nu} \nu}}{\Gamma_{\bar{l} l}}\right)^{SM}
\Gamma_{\bar{l} l}~.
\label{dginv}
\end{equation}
From LEP measurements we have\cite{pdg98}
\begin{equation}
\Gamma_{invis} = 500.1 \pm 1.8~\mathrm{MeV},~~~~~
\Gamma_{\bar{l} l}=83.91 \pm 0.1~\mathrm{MeV},
\label{lepdata}
\end{equation}
then, using the ratio of the neutrino and charged leptons partial widths
calculated within the SM
\begin{equation}
\left(\frac{\Gamma_{\bar{\nu} \nu}}{\Gamma_{\bar{l} l}}\right)^{SM}
= 1.991 \pm 0.001~.
\label{smratio}
\end{equation}
we obtain from \eq{dginv}
\begin{equation}
{\Delta \Gamma_{invis}} \simeq {-1.1} \pm {1.9}~\mathrm{MeV}~.
\label{deltainvexp}
\end{equation}

Therefore, from \eq{widths}, \eq{deltainv} and \eq{deltainvexp} we obtain
for $\bar{b}^2=1$ (pure vector or pure axial $\nu$--$\nu$ interaction)
\begin{equation}
\bar{F} < {90}\, G_F\hspace{0.4cm}  \ ,
\label{vecbound}
\end{equation}
while for
{pure $V-A$} couplings ( $\bar{b}^2=0$) we obtain
\begin{equation}
\bar{F} < {160}\, G_F\hspace{0.4cm} \ .
\label{v-abound}
\end{equation}

The upper bound on this constant is much lower than earlier existing 
particle physics bounds and one order of magnitude lower
than the estimate obtained from the supernova neutrino diffusion time
\footnote{Note also the important improvement with respect the results obtained
in \cite{BBS93} which is due to the updated values of LEP results we 
use here.}.

\section{{$W^+ \rightarrow \ell^+ \nu \nu \bar\nu$}}
\label{sec:wdecay}

Four-neutrino interactions will also give rise to the decay
\begin{equation}
W^+\rightarrow \ell^+ \nu_\ell \nu \bar{\nu}
\label{wdecay}
\end{equation}
Using our effective Hamiltonian for the SNI we get the following
lepton spectrum in the rest frame of the $W$.
\[
\frac{d\Gamma}{dE}=\frac{1}{9}\frac{1}{(2\pi)^5}\frac{G_F}{\sqrt{2}}
F^2 m_W^2 a_\ell^2\left[a_\ell^2+
\sum_{\ell'}(a_{\ell'}^2+b_{\ell'}^2)\right]
\]
\begin{equation}
\times\sqrt{E^2-m_\ell^2} (E_0-E)(3m_W E-2E^2-m_\ell^2)
\label{wspectrum} \ ,
\end{equation}
where E is the total energy of the charged lepton, $m_W$ and $m_\ell$ are
the masses of the $W$-boson and the lepton and $E_0=(m_W^2+m_\ell^2)/2m_W$ is
the maximum energy of the lepton.

The search for {decays of the $W$-boson} with a 
{single lepton with energy less than $E_0$} could give additional 
information about $\nu-\nu$ interactions.
This analysis could be done using present LEPII data.

\section{``Secret'' neutrino interactions at one loop}
\label{sec:2nus}

All previous bounds are extracted from processes in which the new interaction 
is the only relevant one and, therefore, observables 
depend quadratically on the coupling $F$. Obviously, if the new interaction 
enters in loop corrections
to a SM process,  modifications come through its interference
with the SM amplitude and, then, the deviations from the SM predictions
will depend linearly on the coupling $F$.

For example, $\nu$--$\nu$ interactions will
contribute to the decay {$Z \rightarrow \bar{\nu} \nu$} at 
the one-loop level and consequently to the invisible width of the 
$Z$-boson\cite{bs94}.

It is very simple to estimate the order of magnitude of
the corresponding contribution of the non-standard interactions 
at one-loop:
\begin{equation}
\frac{\Delta \Gamma_{\bar{\nu} \nu}}{\Gamma_{\bar{\nu} \nu}} \approx
\frac{F M_Z^2}{(4\pi)^2}~.
\end{equation}

\begin{figure}[t]
\begin{center}
\epsfxsize=10pc 
\epsfbox{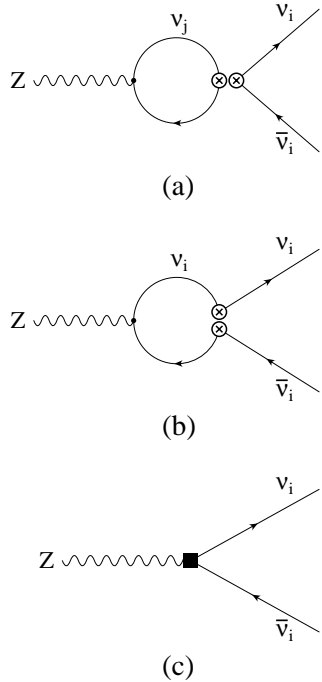}
\end{center} 
\caption{Diagrams that give contributions
to the $Z\bar{\nu} \nu$ vertex in the presence of the non-standard
four-neutrino interaction. In diagram (a), neutrinos of different
flavours are running in the loop.\label{fig:nunuloop}}
\end{figure}

As the invisible width of the $Z$-boson is now measured with an accuracy
better than 1\%, one finds the following bound
on the non-standard coupling $F$:
\begin{equation}
F \le (1\mathrm{-}10)\cdot G_F.
\label{estimate}
\end{equation}
Therefore, one expects stronger bounds of $F$ coming from
the one-loop analysis than those which follow from its contribution to
the invisible width of the $Z$-boson at tree-level.
\par
Although four-fermion interactions are not renormalizable in the 
``text-book-sense'', still one can obtain some information on their couplings 
by using them at the one-loop level\cite{bs94,EQFT}. This is done by 
considering additional 
dimension-six operator(s) which mix with the four-fermion
operator under the renormalization group and serve as counterterms to cancel
the divergences arising from the use of the four-fermion operator in the loop.
The price to be paid is the introduction of more unknown parameters in
the analysis which depend on the details of the full theory giving rise to 
the effective theory. However, if the scale of new physics and the EW scale 
are well separated, the dominant contributions are the logarithmic terms
coming from the running between the two scales. These contributions are
quite model independent and can be unambiguously computed in the effective 
theory.

The partial decay width of the $Z$-boson
into two neutrinos can be written in the following form:
\begin{equation}
\Gamma(Z\rightarrow \bar{\nu} \nu) =
\Gamma^{SM}(Z\rightarrow \bar{\nu} \nu)
+\Delta \Gamma_{\bar{\nu} \nu}~,
\label{gamnunu}
\end{equation}
where $\Gamma^{SM}(Z\rightarrow \bar{\nu} \nu)$
is the SM contribution
and $\Delta \Gamma_{\bar{\nu}\nu}$ contains the effects of the non-standard
operators.

At lowest order these effects come from the interference of the
non-standard amplitude with the SM amplitude and we have
\begin{equation}
{\Delta \Gamma_{\bar{\nu} \nu}}~=~
\Gamma^{SM}(Z\rightarrow \bar{\nu} \nu) 2 {\Re{g_L(M_Z^2)}},
\label{dgamnunu}
\end{equation}
where 
\begin{equation}
\Re{g_L(q^2)} = G_F {q^2} \left(c_2+c_1 \kappa+c_1 \gamma
{\log(M_Z^2/|q^2|)}
\right)~,
\label{regl}
\end{equation}
gives the vertex $Z \nu \bar\nu$ induced by the four-fermion interaction
at one loop. The constants $\gamma$ and $\kappa$ and $c_1$ are\cite{bs94}
\begin{equation}
\gamma=\frac{1}{3\pi^2}~,~~~~
\kappa=\frac{17}{36\pi^2}~,~~~~
c_1 = \frac{{F} a^2 }{G_F} = \frac{\bar{F}}{G_F} ~,
\end{equation}
and {$c_2$} is just the finite part of the counterterm needed 
to absorb the divergences encountered in the loop calculation
(see fig.~\ref{fig:nunuloop}c).

Because the standard $Z\bar{\nu}\nu$ coupling
only involves left-handed neutrinos, and
because the lowest order effect of the non-standard interaction comes
via its interference with the standard coupling,
only interactions of left-handed neutrinos contribute. 

Assuming that there are three generations of neutrinos, the non-standard
contribution to the invisible width of the $Z$-boson is now
\begin{equation}
\Delta \Gamma_{invis}=3\Delta \Gamma_{\bar{\nu} \nu}~,
\label{deltainv2nu}
\end{equation}
where $\Delta \Gamma_{\bar{\nu} \nu}$ is given above.
Using the limits on $\Delta \Gamma_{invis}$ obtained in section \ref{sec:4nus}
we get 
\begin{equation}
\label{nnbounds}
-0.03 \le c_2 +c_1 \kappa \le 0.007~.
\end{equation}
If there are no unnatural cancellations between the couplings
each of the couplings can be bounded independently of the others and we 
obtain:
\begin{equation}
\abs{c_1} = \frac{{\bar{F}}}{G_F} \le {0.6},\ \ \ 
\abs{c_2} \le 0.039,\ \ \ 
\end{equation}

However, even if there are cancellations
at this particular scale ($q^2 = m^2_Z$) there 
will not be cancellations at
other scales, because the logarithmic dependence of $g_L(q^2)$ on
$q^2$.  Therefore, it will still be possible to get some interesting
bounds on the coupling $\bar F$ if additional data obtained at different
scales are used\cite{bs94} (for instance DIS experiments at high 
energy $(-q^2\simeq 100-1000~\mathrm{GeV}^2)$.

Thus, from the analysis of LEP data we can say that
contact four-fermion neutrino interactions, involving only left-handed
neutrinos, cannot be larger than standard neutral current interactions.

\section{{Primordial Nucleosynthesis}}
\label{sec:ns}

Data on primordial nucleosynthesis offers a limit on the number of the
massless degrees of freedom contributing to the early universe expansion 
for temperatures $T \ge 1$~MeV. 

Three right-handed neutrinos in equilibrium
with their left-handed partners
at these temperatures are completely excluded. In fact, the three
right-handed neutrinos should have decoupled at $T \simeq 200$~MeV.

Four-neutrino interactions of the type considered involving both,
left-handed and right-handed neutrinos, could keep right-handed neutrinos
in thermal equilibrium\cite{mt94} through the reactions
\begin{equation}
\bar{\nu}_{Li} + \nu_{Ri} \Leftrightarrow \bar{\nu}_{Rj} + \nu_{Lj}
\end{equation}

Requiring that there is decoupling at a temperature $T$, that is, enforcing
that the interaction rate $\Gamma$ is smaller that the expansion rate of
the universe $H$, one obtains a rather stringent limit on four-neutrino
couplings\cite{mt94}:
\begin{equation}
{F_V} < {3\times 10^{-3}} G_F
\end{equation}
for pure vector interactions. 

If four-fermion interactions involve only left-handed 
(or only right-handed) neutrinos the limit
does not apply at all.

Other constraints can be obtained from ultra-high energy AGN 
neutrinos\cite{keranen}.
However they are only relevant for relatively light mediators
($m_V < 0.5$~GeV). 

\section{{Conclusions}}
\label{sec:conclusions}

We have reviewed, from a historical point, the information obtained on
the possible existence of a strong four-fermion contact neutrino interaction.

Bounds on four-neutrino interactions coming from {$K^+$ and $\pi^+$}
decays are very soft and still allow for interactions much stronger than
standard model interactions. 

From neutrino diffusion time in the
supernova one can set better bounds but ``Secret'' interactions could 
still be large.

The invisible decay width of the $Z$ gauge boson constrains 
strongly four-neutrino interactions, 
if they involve left-handed neutrinos. It gives contributions to four-neutrino
decay at tree level and to two neutrino decay at one-loop. Using present data
we find that SNI interactions involving left-handed neutrinos 
cannot be larger than neutral current standard model interactions.

``secret'' interactions involving both, left-handed and right-handed
neutrinos are severely constrained by primordial nucleosynthesis. 
Their strength must be at least two orders of magnitude smaller than in the
standard model, although, these bounds do not apply to pure
left-handed or pure right-handed couplings. 

Taking all information together we find that there is no room for 
strong four-fermion contact neutrino interactions unless 
they involve {\em only} right-handed neutrinos.

\section*{Acknowledgements}

We would like to acknowledge especially Samoil Bilenky for an enjoyable 
collaboration and for introducing us to the subject discussed in this 
paper. Many of the results presented here belong to Samoil. A.S. also likes
to thank Wanda Alberico for giving him the opportunity to join Samoil and 
his friends in this special occasion. Finally A.S. is also indebted to
the CERN Theory division, where this work has been written in part, for its 
hospitality.
This work has been funded, in part, by CICYT under the Grant AEN-96-1718, 
by DGESIC under the Grant PB97-1261 and by the DGEUI of 
the Generalitat Valenciana under the Grant GV98-01-80.


\begin{thebibliography}{99}

\bibitem{bile1} D. Yu. Bardin, S.M. Bilenky and B. Pontecorvo,
\Journal{\em Phys. Lett.}{32B}{121}{1970}.

\bibitem{bbnc} Z. Bialynicki-Birula,
\Journal{\em Nuovo Cimento}{33}{1484}{1964}. 

\bibitem{singlet}  Y. Chikashige, R.N. Mohapatra and R.D. Peccei,
\Journal{\em Phys. Lett.}{98B}{265}{1981}.

\bibitem{noytriplet}
A.~Santamaria, \Journal{\PRD}{39}{2715}{1989}.


\bibitem{triplet} G.B. Gelmini and M. Roncadelli,
\Journal{\em Phys. Lett.}{99B}{441}{1981}; H.M. Georgi, S.L. Glashow and
S. Nussinov, \Journal{\NPB}{193}{297}(1981).

\bibitem{doublet} S. Bertolini and A. Santamaria,
\Journal{\NPB}{310}{714}{1988}.

\bibitem{susymajoron}
A.~Santamaria and J.W.~Valle,
\Journal{\em Phys. Lett.}{195B}{423}{1987};
G.G.~Ross and J.W.~Valle,
\Journal{\em Phys. Lett.}{151B}{375}{1985};
C.S.~Aulakh and R.N.~Mohapatra,
\Journal{\em Phys. Lett.}{121B}{147}{1983}.

\bibitem{majbounds} V. Barger, W.Y. Keung and S. Pakvasa,
\Journal{\PRD}{25}{907}{1982}.

\bibitem{CHP}
{G.D. Cable, R.H. Hildebrand, C.Y. Pang and R. Stiening},
\Journal{\em Phys. Lett.}{40B}{699}{1972};
{C.Y.~Pang, R.H.~Hildebrand, G.D. Cable and R. Stiening},
\Journal{\PRD}{8}{1989}{1973}.

\bibitem{final} R.S. Hayano {\it et al}, \Journal{\PRL}{49}{1305}{1982}; 
D.I. Britton {\it et al}, \Journal{\PRD}{46}{R885}{1992} 

\bibitem{sn87back} E.W. Kolb, M.S. Turner, 
\Journal{\PRD}{36}{2895}{1987}.

\bibitem{sn87diff} A. Manohar, 
\Journal{\PLB}{192}{217}{1987}.

\bibitem{Dicus:1989jh}
D.A.~Dicus, S.~Nussinov, P.B.~Pal and V.L.~Teplitz,
\Journal{\PLB}{218}{84}{1989}.

\bibitem{BBS93}
{M. Bilenky, S.M. Bilenky and A. Santamaria},
\Journal{\PLB}{301}{287}{1993}.

\bibitem{pdg98} Particle Data Group: C.~Caso {\em et al.}, 
\Journal{{\em Eur. Phys. J.} C}{3}{1}{1998}.

\bibitem{bs94} M. Bilenky and A. Santamaria, 
\Journal{\PLB}{336}{91}{1994}.

\bibitem{EQFT}
For review of the use of EQFT, see e.g.: H. Georgi,
\Journal{\em Annu. Rev. Nucl. Sci.}{43}{209}{1993}.

\bibitem{mt94} E. Mass\'o and R. Toldr\`a, 
\Journal{\PLB}{333}{132}{1994}.

\bibitem{keranen} P.~Ker\"anen, 
\Journal{\PLB}{417}{320}{1998}.

\end{thebibliography}
\end{document}